\documentclass[global,twocolumn]{svjour}
\usepackage{graphicx}
\journalname{Applied Physics B}
\makeatletter
\def\lambdabar{\protect\@lambdabar}
\def\@lambdabar{%
\relax
\bgroup
\def\@tempa{\hbox{\raise.73\ht0
\hbox to0pt{\kern.25\wd0\vrule width.5\wd0
height.1pt depth.1pt\hss}\box0}}%
\mathchoice{\setbox0\hbox{$\displaystyle\lambda$}\@tempa}%
{\setbox0\hbox{$\textstyle\lambda$}\@tempa}%
{\setbox0\hbox{$\scriptstyle\lambda$}\@tempa}%
{\setbox0\hbox{$\scriptscriptstyle\lambda$}\@tempa}%
\egroup
}
\makeatother

\begin{document}

\title{Coherent transport of matter waves}

\author{Carsten Henkel$^1$ 
\and Sierk P\"otting$^{2,3,4}$}

\institute{$^1$Institut f\"ur Physik, 
Am Neuen Palais~10, Universit\"at Potsdam,
14469 Potsdam, Germany,  
\email{Carsten.Henkel@quantum.physik.uni-potsdam.de} 
\\
$^2$Optical Sciences Center, University of Arizona, Tucson, AZ 85721,
USA \\
$^3$Max--Planck--Institut f\"ur Quantenoptik, 85748 Garching, Germany \\
$^4$Sektion Physik, Universit\"at M\"unchen, 80333 M\"unchen, Germany}
\date{17 July 2000}

\titlerunning{Coherent transport of matter waves}
\authorrunning{C. Henkel and S. P\"otting}

\maketitle

\abstract{
A transport theory for atomic matter waves in low-dimensional waveguides is outlined. 
The thermal fluctuation spectrum of magnetic near fields leaking out of metallic
microstructures is estimated. The corresponding scattering rate for paramagnetic atoms 
turns out to be quite large in micrometer-sized waveguides (approx.\ 100/s). 
Analytical estimates for the heating and decoherence of a cold atom cloud are given. 
We finally discuss numerical and analytical results for the scattering from static potential 
imperfections and the ensuing spatial diffusion process.
\vspace*{2mm}
}



\section*{Introduction}

Atom optics, the coherent manipulation of atomic matter waves, is currently
heading towards miniaturization and integration. Atom guides in
both one and two dimensions have been demonstrated 
\cite{Schmiedmayer95,Ohtsu96b,Mlynek98b,Schmiedmayer98b,Hinds00}.
Recent experiments have studied atoms trapped in electromagnetic solid-state 
hybrid surface guides (or ``atom chips'') 
\cite{Anderson99,Schmiedmayer00a,Prentiss00,Schmiedmayer00b}.
The near future may be expected to see the trapping and manipulation 
of Bose-condensed atoms in such microtraps \cite{Grimm00a}.
Decoherence is an intriguing issue in this context because of 
the close proximity of the cold atom cloud to the macroscopic substrate, 
typically being held at room temperature. Thermal electromagnetic fields
thus perturb the atoms, leading to heating, trap loss and scattering
\cite{Henkel99b,Henkel99c}. The thermal noise spectra are much larger than
the blackbody spectrum because the characteristic distances
are much smaller than the photon wavelengths for the relevant transition
frequencies: the atoms are subject to thermal near fields
\cite{Henkel00b}. It is therefore of much interest to estimate the length
and time scales over which the transport of matter waves in atom chips
remains coherent.

In this paper, we outline a transport theory for matter waves in 
low-dimensional waveguides close to metallic microstructures. 
Such structures are used in current experiments to generate
magnetic fields or to reflect optical fields. We focus on paramagnetic atoms 
and their perturbation by the thermally excited magnetic near field. 
Scattering rates and noise spectra are computed for a few generic geometries: 
a metallic half-space, a layer, and a cylindrical wire. A self-consistent
transport theory for the atomic Wigner distribution is formulated for
a dilute, non-condensed cloud. In the case of a white noise spectrum 
(that turns out to be an excellent approximation for the magnetic near
field), the transport equation is solved analytically, and the atomic
decoherence rate is determined. We finally present analytical and numerical
calculations for a waveguide with static roughness (due to imperfections
in the guiding field, for example).

\section{Model}

\subsection{Perturbation of a trapped spin}

We start from an atom trapped in a wave guide potential
that restricts its motion to one or two dimensions. 
We suppose that the transverse degrees of freedom are `frozen out',
the atom being in the transverse ground state. In the remaining
directions, we assume a free motion. Examples of such waveguides
are linear magnetic quadrupole guides formed by currents in one or
several parallel wires 
\cite{Schmiedmayer95,Anderson99,Schmiedmayer00a,Prentiss00,Schmiedmayer00b}
or planar guides mounted above a surface `coated' with repulsive 
optical \cite{Mlynek98b} or magnetic \cite{Hinds00} fields.
Trapped paramagnetic atoms couple to thermal fluctuations of the
magnetic field and will scatter from these at a rate given by
Fermi's Golden Rule
\begin{equation}
\gamma_{\rm i\to f} =
\frac{ 1 }{ \hbar^2 }
\sum_{i,j = 1}^{3}
\langle {\rm i} | \mu_i | {\rm f} \rangle
\langle {\rm f} | \mu_j | {\rm i} \rangle
B_{ij}( \omega_{\rm if} )
,
\label{eq:flip-rate}
\end{equation}
where $B_{ij}( \omega_{\rm if} )$ is the magnetic
cross correlation tensor (spectral density) 
and $\hbar\omega_{\rm if}$ is the energy difference 
between the initial and final states $|{\rm i}\rangle, \, |{\rm f}\rangle$. 
The correlation tensor $B_{ij}( \omega )$
is calculated in subsection~\ref{s:magnetic-near-field}, where we shall
see that it depends on the distance to the metallic microstructures.
Its frequency dependence is quite weak, however, because the relevant
transition frequencies are at most in the GHz range. Magnetic noise
at these frequencies drives spin-flip transitions that, in a magnetostatic
trap, put the atom onto a non-trapping potential surface. The concomitant
loss rate has been discussed previously \cite{Henkel99c}. We focus here
on the scattering between quasi-plane waves in the same spin state
that remain trapped in the waveguide. These transitions  
correspond to much lower frequencies of the order of the temperature
of the atomic cloud (typically in the kHz range). The corresponding
photon wavelengths are much larger than the characteristic size of the
waveguide structures. For this reason, we may adopt the magnetostatic 
approximation in order to compute the thermal magnetic field spectral 
density.

\subsection{Magnetic near field}
\label{s:magnetic-near-field}
\subsubsection{The method.}

Fluctuation electrodynamics (or source theory) is a convenient tool
to compute the spectral correlation tensor of the electromagnetic field
in an inhomogeneous, dissipative environment. Basically, one introduces 
a fluctuating current density whose spectral density 
is determined from the imaginary 
part of the dielectric function. The field radiated by these currents
is computed from the Green tensor for the chosen geometry
\cite{Lifshitz56,Varpula84}. It has been
shown that this scheme also provides a consistent way to quantize the
electromagnetic field in dispersive and absorbing dielectrics 
\cite{Welsch96,Knoell98a,Knoell98b,Savasta00} 
(see \cite{Barnett97} for further reference). 

\subsubsection{Low-frequency noise.}

For our present purposes, a simplified Green tensor is sufficient:
we focus on nonmagnetic media and low frequencies where the magnetostatic
approximation is valid. As outlined in appendix~\ref{a:noise-currents},
we find the following expression for the scattering rate~(\ref{eq:flip-rate})
\begin{equation}
\gamma = C_0 Y_{nn}
,
\label{eq:general-estimate}
\end{equation}
where the prefactor is
\begin{equation}
C_0 = 
\frac{ \left| \langle s | \mu_n | s \rangle \right|^2 }{ \hbar^2 }
\frac{ k_{\rm B} T }{ 4\pi^2 \varepsilon_0^2 c^4 \varrho }
.
\label{eq:low-freq-limit}
\end{equation}
Here, $\mu_n$ is the projection of the magnetic moment along a static bias
field, $|s\rangle$ is the trapped spin state, and $T$, $\varrho$ are 
temperature and specific resistance of the metallic microstructures.
Finally, $Y_{nn}$ is the matrix element along the bias field of a
geometric tensor $Y_{ij}$ defined in~(\ref{eq:X-tensor}). The scattering 
rate~(\ref{eq:general-estimate}) is of the order of
\begin{equation}
\gamma \sim 
75\,{\rm s}^{-1} 
\frac{ ( \mu / \mu_{\rm B} )^2 (T / 300\,{\rm K}) }
{ ( \varrho / \varrho_{\rm Cu} ) }
( Y_{ij}\times 1\,\mu{\rm m} )
.
\label{eq:estimate-gamma}
\end{equation}
(We have taken the specific resistance of copper $\varrho_{\rm Cu} = 1.7 \times 
10^{-6}\,{\rm\Omega\,cm}$ \cite{Kittel}.)
The geometric tensor $Y_{ij}$~(\ref{eq:X-tensor}) has dimension (1/length),
and its magnitude is the inverse of the characteristic scale of the waveguide.
As a consequence, the scattering rate~(\ref{eq:estimate-gamma}) is quite 
large for a typical micrometer-size waveguide.

\subsubsection{Typical geometries.}

We now give the results of the evaluation of the geometric 
tensor~(\ref{eq:X-tensor})
for three generic geometries. For a metallic half-space, we find that
\begin{equation}
Y_{ij}
=
\frac{ \pi t_{ij} }{ 4 z } 
,
\label{eq:half-space}
\end{equation}
where the diagonal tensor $t_{ij}$ has elements $(\frac32, \frac32, 1)$.
Note the long range $1/z$ dependence on distance $z$ 
\cite{Henkel99b}.

For a metallic layer of thickness $d$ above a nonconducting substrate, we 
find
\begin{equation}
Y_{ij}
=
\frac{ \pi t_{ij} d }{ 4 z ( z + d ) } 
.
\label{eq:layer}
\end{equation}
This expression reproduces the result~(\ref{eq:half-space}) for the 
metallic half space when the distance $z$ from the upper interface 
is much smaller than the layer thickness $d$. On the other hand, 
at large distances a faster decrease $d/z^2$ takes over.

\begin{figure}
\centerline{%
\resizebox{0.7\columnwidth}{!}{%
\includegraphics*{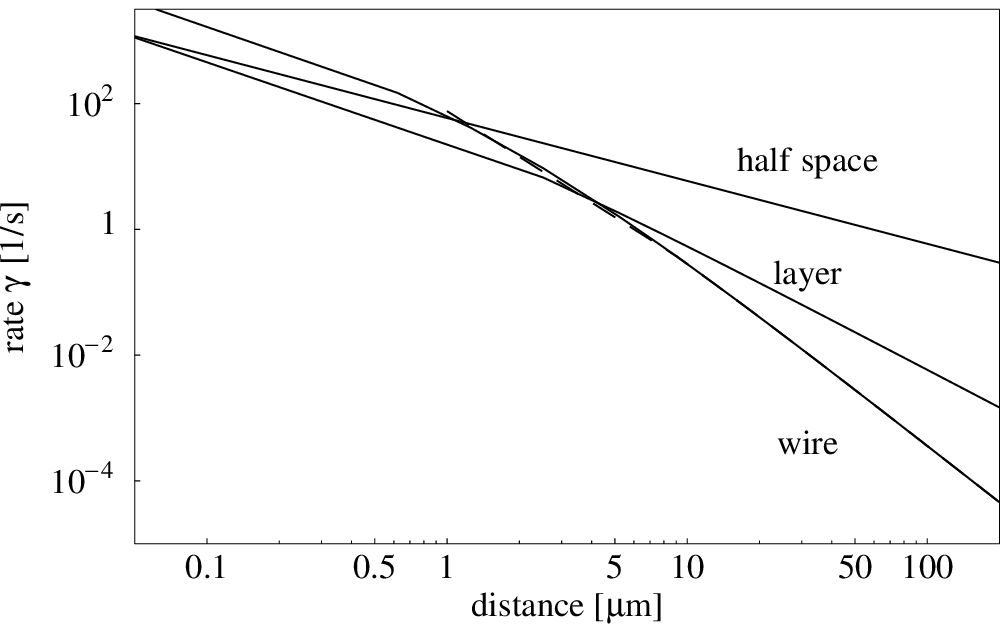}}}
\caption[]{Scattering rate in a waveguide at distance $z$ from three
metallic structures: half-space, thin layer, thin wire. The dashed
line gives the large-distance limit~(\ref{eq:wire-far}) for the wire.
The metal is copper at room temperature. 
Both the layer and the wire have a thickness $1\,\mu$m.}
\label{fig:rates}
\end{figure}

Finally, for a cylindrical wire of radius $a$, we found a cumbersome expression
involving an elliptic integral. Nevertheless, simple results are obtained
for the trace of the geometry tensor in the following limits:
(1) If the distance $R$ from the wire axis is large compared to wire radius, 
\begin{eqnarray}
&&
{\rm tr}\,Y_{ij}
\approx
\frac{ \pi^2 a^2}{ 2 R^3 }
\left[
1
+
\frac 94
\left( \frac a R \right)^2
+ 
{}\right.
\nonumber\\
&&
\qquad\left.{}+
\frac{ 225 }{ 186 }
\left( \frac a R \right)^4
+
{\cal O}[ (a/R)^6 ]
\right]
.
\label{eq:wire-far}
\end{eqnarray}
Note the even stronger power law decrease $a^2/R^3$ to leading order
compared to the planar geometries. 
The three-term expansion~(\ref{eq:wire-far}) is reasonably accurate 
down to $R \ge 1.6\,a$. (2) In the short-distance limit $R - a \ll a$, 
we recover the flat 
half-space result~(\ref{eq:half-space})
\begin{equation}
{\rm tr}\,Y_{ij}
\approx
\frac{ \pi }{ R - a }
.
\label{eq:wire-close}
\end{equation}
In Figure~\ref{fig:rates}, we plot the corresponding scattering rates
as a function of distance (i.e., $z$ for the planar guides, $R-a$ for the wire
guide). The static bias field at the waveguide center is taken perpendicular 
to the metallic surface. The scattering rate for the wire is actually 
overestimated because the trace of the magnetic correlation
tensor is used.

\subsection{Transport equation}

The evolution of a trapped matter wave in the waveguide is conveniently
described in terms of a transport equation. This equation allows to
characterize the evolution of the single-particle spatial
density matrix (or coherence function)
\begin{equation}
\rho( {\bf x}; {\bf s} ) = 
\langle
\psi^*( {\bf x} + {\textstyle\frac12} {\bf s} )
\,
\psi( {\bf x} - {\textstyle\frac12} {\bf s} )
\rangle
,
\label{eq:coh-function}
\end{equation}
where the average $\langle \ldots \rangle$ is taken over the spatial
and temporal fluctuations of a perturbing potential. Instead of working
with the coherence function~(\ref{eq:coh-function}), we formulate the
transport equation for the Wigner transform of the density matrix
\begin{equation}
W( {\bf x}, {\bf p} ) = \int \!
\frac{ d^Ds }{ (2 \pi\hbar)^D } e^{ i {\bf p} \cdot {\bf s} / \hbar }
\rho( {\bf x}; {\bf s} )
,
\label{eq:def-Wigner}
\end{equation}
that may be interpreted as a quasi-probability distribution in phase
space. Here and in the following, the coordinates ${\bf x}$ and ${\bf p}$ 
describe the motion in the $D = 1,\,2$ dimensional waveguide. 

The quasi-free wave function for the trapped spin state $|s\rangle$ in
the waveguide is perturbed by the magnetic potential
\begin{equation}
V( {\bf x}, t ) =  \langle s | \mu_n | s \rangle B_n( {\bf x}, t )
,
\label{eq:perturbation}
\end{equation}
where $B_n$ is again the field component along the static bias field.
We now have to take into account both energy and momentum changes 
in the scattering process. This may be done in terms of a master (or
transport equation) for the Wigner distribution. It is derived
using second order perturbation theory with respect to $V$, assuming 
that $V$ has gaussian statistics and doing a multiple-scale
expansion of the Bethe-Salpeter equation for the coherence function
\cite{Keller96}. We find:
\begin{eqnarray}
&&
\Bigl(
\partial_t + 
\frac1m {\bf p} \cdot \nabla_{\bf x} 
+ 
{\bf F}_{\rm ext} \cdot \nabla_{\bf p}
\Bigr)
W( {\bf x}, {\bf p} )
=
\label{eq:transport-eq}\\
&& 
\int\!{\rm d}^Dp' \,
S_V( {\bf p} - {\bf p}'; E_{p} - E_{p'} )
\left[
W( {\bf x}, {\bf p}' )
-
W( {\bf x}, {\bf p} )
\right]
,
\nonumber
\end{eqnarray}
with the de Broglie dispersion relation $E_p = p^2/2m$ and where
$S( {\bf q}; \Delta E)$ is the spectral density of the perturbation
\begin{eqnarray}
&&
S_V( {\bf q}; \Delta E) =
\frac{ 1 }{ \hbar^2 }
\int\!\frac{ d^D s \, d\tau}{ (2\pi\hbar)^{D} }
\langle
V( {\bf x} + {\bf s}, t + \tau) \,
V( {\bf x}, t ) 
\rangle
\times{}
\nonumber
\\
&&
\qquad
{} \times
\, e^{ - i ( {\bf q}\cdot{\bf s} - \Delta E \tau ) / \hbar}
.
\label{eq:potential-spectrum}
\end{eqnarray}
The average $\langle \ldots \rangle$ is again taken with respect to the
magnetic noise field. We assume for simplicity that the noise is statistically 
stationary in time and along the waveguide directions. This is
actually the case for linear or planar waveguides parallel to 
planar structures and for a linear guide parallel to a wire. 

The left hand side of the transport equation~(\ref{eq:transport-eq})
describes the ballistic motion of the atom subject to the external 
(deterministic) force ${\bf F}_{\rm ext}$. The right hand side
describes the scattering off the magnetic field fluctuations. 
As a function of the momentum transfer $\bf q$, \emph{e.g.},  
the spectral density $S_V( {\bf q}; \Delta E)$ is proportional to the 
spatial Fourier transform of the potential, as to be expected from the Born 
approximation for the scattering process ${\bf p}' \to {\bf p} =
{\bf p}' + {\bf q}$. The transport equation thus combines in a 
self-consistent way ballistic motion and scattering processes.

\begin{figure}
\centerline{%
\resizebox{8.5cm}{!}{%
   \includegraphics*{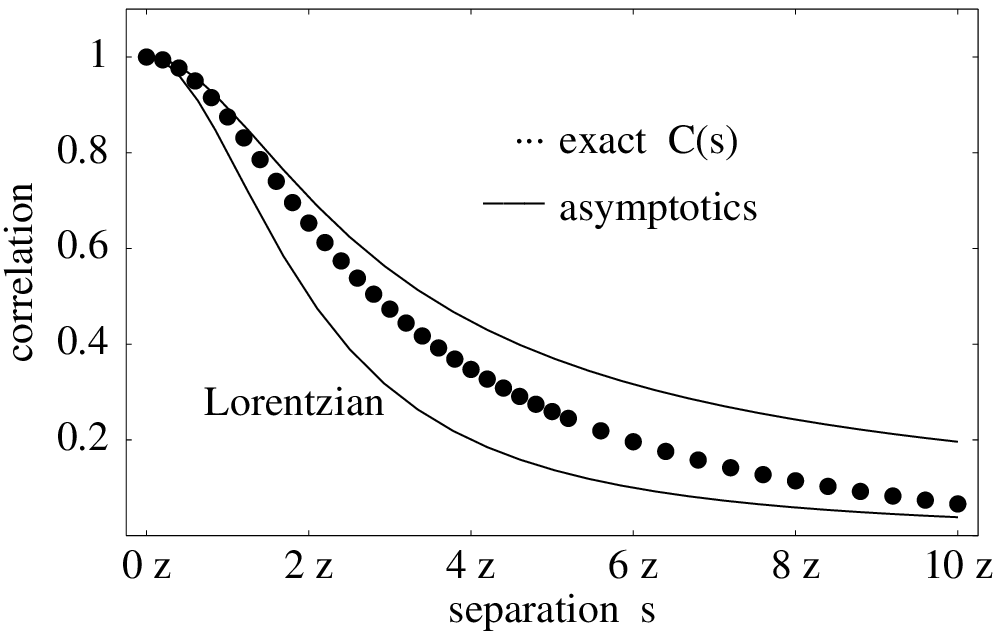}
   \hspace*{10mm}
   \includegraphics*{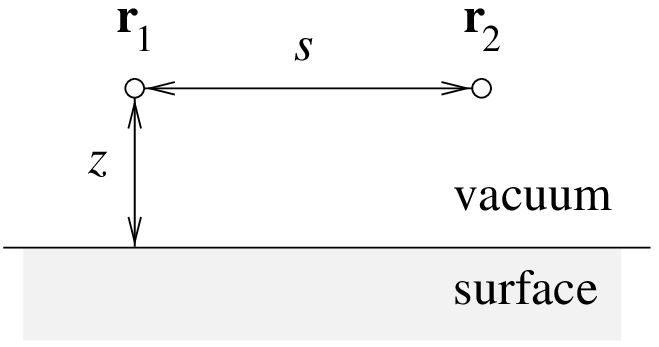}
   }}
\caption[]{%
Spatial (normalized) correlation function of the thermal magnetic near 
field above a metallic surface at frequency $\omega/2\pi = 30\,$MHz
(copper at 300~K). 
The separation $s$ gives the distance between two observations points 
at the same height $z$ above the surface.
Dots: exact evaluation, solid lines: asymptotic expansions in the
short-distance regime, discussed in~\cite{Henkel00b}.
For lower frequencies, the
correlation function remains essentially unchanged. 
}
\label{fig:lateral-corr}
\end{figure}

The frequency dependence of the magnetic field noise has already been
treated in the previous subsection. We saw that the spectral density 
is flat in
the magnetostatic approximation. The magnetic perturbation can therefore 
be treated as a white noise. To get the wave vector dependence, we have
to evaluate the spatial Fourier transform of the two-point correlation function
$B_{ij}( {\bf x}_1, {\bf x}_2 )$. In the magnetostatic limit, it turns out
that above a planar metallic layer, the correlation function is similar
to a lorentzian, with a correlation length $l_c$ parallel to the layer 
given by the distance $z$ perpendicular to the layer (see 
figure~\ref{fig:lateral-corr}). More details are discussed elsewhere
\cite{Henkel00b}. As a consequence of the finite correlation length, 
the scattering kernel $S( {\bf q}; \Delta E)$ in the transport 
equation~(\ref{eq:transport-eq}) exhibits a cutoff for momentum
transfers $|{\bf q}| \ge \hbar / l_c \approx \hbar / z$. The spatial
smoothness of the magnetic field thus limits the possible scattering 
processes, although arbitrary energy transfers are available from the 
white noise perturbation.

In the following, we assume a constant force ${\bf F}_{\rm ext}$
and focus on two limiting cases of the transport
equation~(\ref{eq:transport-eq}): (1) `Inelastic transport'
under the influence of white noise perturbations. This is the appropriate
limit to estimate the detrimental effects of thermal magnetic fields.
(2) `Elastic transport' where a time-independent perturbation is assumed.
This allows to describe in the same framework the influence of static
roughness in the waveguide potentials from which the atoms scatter.
In optical potentials, roughness of this type occurs when the optical
fields are scattered from small-scale inhomogeneities in the microstructure
\cite{Henkel97a}. In planar gravito-optical traps based on a planar 
evanescent wave mirror, the rough optical potential couples efficiently 
the atomic motion normal and parallel to the mirror \cite{Grimm97b}. 
The transport equation is valid as long as one is interested in the 
evolution on length scales large compared to the correlation length of 
the roughness. In rough evanescent fields, we find again that this length 
is typically of order $z$ \cite{Henkel97a} because high spatial 
frequencies $K \gg 2\pi/\lambda$ give rise to exponentially damped 
fields $\propto \exp(- K z)$.

\section{Results}

\subsection{Inelastic transport}

\subsubsection{Analytic solution of the transport equation.}

For a broad band spectrum of the perturbation, we may neglect the
dependence of $S_V( {\bf q}, \Delta E )$ on $\Delta E$. The transport
equation~(\ref{eq:transport-eq}) then simplifies because the integration 
over the initial momentum ${\bf p}'$ is not restricted by energy
conservation. Taking the Fourier transform of the Wigner function
$W({\bf x}, {\bf p})$ with respect to both
variables ${\bf x}$ and ${\bf p}$ (with conjugate variables ${\bf k}$
and ${\bf s}$), it is simple to derive the following solution
\begin{eqnarray}
&& \tilde{W}( {\bf k}, {\bf s}; t ) =
\tilde{W}_0( {\bf k}, {\bf s} - \hbar{\bf k}t/m ) 
\,
{\rm e}^{ - {\rm i} {\bf F}_{\rm ext} \cdot {\bf s} t / \hbar }
\times {}
\nonumber\\
&& {} \times  
\exp{\left[
- \gamma \int_0^t\!\!
\left(
1 - C( {\bf s} - \hbar{\bf k} t' /m )
\right)\!{\rm d}t'
\right]
}
.
\label{eq:analytic-solution}
\end{eqnarray}
Here, $\tilde{W}_0( {\bf k}, {\bf s})$ is the double Fourier transform
of the Wigner function at initial time $t = 0$, 
and $\gamma$ and the normalized spatial correlation function $C({\bf s})$ 
are related to the correlation function of the perturbation by
\begin{equation}
\gamma C( {\bf s} ) =
\frac{ 1 }{ \hbar^2 }
\int_{-\infty}^{+\infty}\!d\tau\,
\langle V( {\bf s}, \tau ) \, V( {\bf 0}, 0 ) \rangle
,\quad
C( {\bf 0} ) = 1. 
\label{eq:decoherence-rate}
\end{equation}
We also note that $\gamma$ is the transition rate $\gamma_{\rm i \to f}$ 
for scattering processes with zero energy transfer ($\omega_{\rm if} = 0$). 
For a wave guide above a metallic surface, the 
estimate~(\ref{eq:estimate-gamma}) is hence applicable for the rate $\gamma$.

From the analytic solution~(\ref{eq:analytic-solution}), it is easily
checked that in the absence of the perturbation, 
the spatial width $\langle \delta x^2(t) \rangle $ of a cloud 
increases ballistically according to $\langle \delta x^2(t)\rangle
 = \langle \delta p^2(0) \rangle \, t^2 / m^2$ where
$\langle \delta p^2(0)\rangle $ is the initial width 
of the cloud in momentum space (this latter width remains 
constant in this case, of course).

\subsubsection{Spatial decoherence.}

\begin{figure}
\centerline{%
\resizebox{0.7\columnwidth}{!}{%
   \includegraphics*{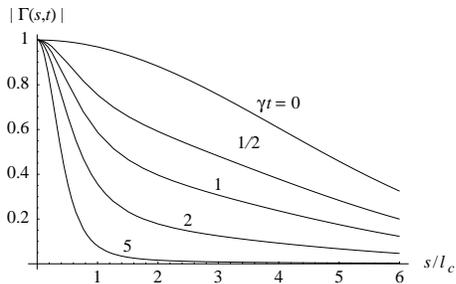}}}
\caption[]{%
Illustration of spatial decoherence in an atomic wave guide. 
The spatially averaged coherence function $\Gamma({\bf s}, t)$
is plotted vs.\ the separation ${\bf s}$ for a few times $t$.
Space is scaled to the field correlation length $l_c$ and time
to the scattering time $1/\gamma$. A Lorentzian correlation
function for the perturbation is assumed.}
\label{fig:inelastic-decoh}
\end{figure}

More interesting information may be obtained for a nonzero scattering 
rate $\gamma$. Note that the spatially averaged atomic coherence function
is given by
\begin{equation}
{\Gamma}( {\bf s} )
= \int\!{\rm d}^D{x} \,
\rho( {\bf x}; {\bf s} )
=
\tilde{W}( {\bf k} = {\bf 0}, {\bf s} )
.
\label{eq:def-coherence-function}
\end{equation}
The solution~(\ref{eq:analytic-solution}) therefore implies that the
spatial coherence decays exponentially with time:
\begin{equation}
{\Gamma}( {\bf s}; t ) = {\Gamma}_0( {\bf s} )
\exp{\Bigl[ 
- 
\gamma t (1 - C( {\bf s} ) ) 
- {\rm i} {\bf F} \cdot {\bf s}\, t / \hbar
\Bigr]}
.
\label{eq:analytic-solution-2}
\end{equation}
The decoherence rate depends on the spatial separation between the points
where the atomic wave function is probed, and is given by 
$\gamma({\bf s} ) = \gamma (1 - C( {\bf s} ) )$. It hence saturates
to the value $\gamma$ at large separations and decreases to zero
for ${\bf s} \to {\bf 0}$. The decay of the coherence 
function~(\ref{eq:analytic-solution-2}) is illustrated in figure~\ref{fig:inelastic-decoh}.
One observes that at time scales $t \ge 1/\gamma$, the spatial 
coherence is reduced to a coherence length $\xi_{\rm coh} \sim l_c$.
After a few collisions with the fluctuating magnetic field, the 
long-scale coherence of the atomic wave function is thus lost and
persists only over scales smaller than the field's correlation length
(where different points of the wave function `see' essentially the
same fluctuations). 
For larger times $t \gg 1/\gamma$, decoherence proceeds at a smaller
rate that is related to momentum diffusion, as we shall see now.

\subsubsection{Momentum diffusion at long times.}

The behaviour of the atomic momentum distribution at long times
may be extracted from an expansion of the analytic 
solution~(\ref{eq:analytic-solution-2}) for small values of ${\bf s}$. Assuming
a quadratic dependence of the field's correlation function,
$C({\bf s}) \approx 1 - s^2/l_c^2$, as one would expect for
Lorentzian correlations, we find that the atomic momentum 
distribution is gaussian at long times; it is centered at ${\bf p}_0 
+ {\bf F}_{\rm ext} t$ due to the external force, and its width
increases according to a diffusion process in momentum space
\begin{equation}
\langle \delta p^2(t) \rangle 
\approx \langle \delta p^2(0) \rangle 
+ \frac{ \hbar^2 \gamma t }{ l_c^2 }
.
\label{eq:p-diffusion}
\end{equation}
This was to be expected: the atoms perform a random walk in momentum
space, exchanging a momentum of order $\hbar/l_c$ per scattering
time $1/\gamma$. The momentum diffusion coefficient $D_p = \hbar^2 \gamma
/ l_c^2$ that may be read off from~(\ref{eq:p-diffusion}) is consistent
with this intuitive interpretation. Physically speaking, the atomic
cloud is `heated up' due to the scattering from the fluctuating potential.
We note that the rate of change of the atomic kinetic energy in the
wave guide plane is the same as the one for the tightly bound motion 
perpendicular to the metallic surface (see \cite{Henkel99c} for a 
calculation of this rate).

Translating the width of the momentum distribution into a spatial coherence 
length, we find a power-law decay at long times, $\xi_{\rm coh} = 
l_c / \sqrt{ \gamma t }$. 
Finally, a similar calculation yields the width of the atomic cloud 
in position space: it increases `super-ballistically' at long times, 
$\langle \delta x^2(t) \rangle \propto t^3$, as a consequence of heating.

\subsection{Elastic transport}

A waveguide potential with static roughness leads to energy-conserving 
scattering. The integration over the initial momenta ${\bf p}'$ in the 
transport equation~(\ref{eq:transport-eq}) is then restricted which
complicates the analytic solution. We discuss in the following 
numerical results and an analytic solution for vanishing external force. 

\subsubsection{Numerical results.}

The transport equation may be solved using a split-step propagation 
method. We alternate between ballistic motion and the scattering.
In one dimension, e.g., the transport equation reads
\begin{eqnarray}
&&
\left( 
\partial_t + \frac pm\partial_x + F_{\rm ext}\partial_p
\right)
W(x,p,t)
=
\\
&&\gamma(p) 
\left(
W(x,-p,t) - W(x,p,t)
\right)
,
\nonumber
\end{eqnarray}
where the scattering rate $\gamma(p)$ depends on momentum via 
the wavevector transfer $|2p/\hbar|$. The solution for pure ballistic 
transport is described by
\begin{eqnarray}
&&
W(x, p, t + \Delta t) = 
\\
&&
W(x - (p/m) \Delta t 
+ (F_{\rm ext}/2m) \Delta t^2, p - F_{\rm ext} \Delta t, t)
,
\nonumber
\end{eqnarray}
while the scattering kernel gives
\begin{eqnarray}
&&
\left(\begin{array}{c} W(x,p,t + \Delta t) \\ W(x,-p,t + \Delta t) \end{array}\right)
=
\label{eq:scattering-step}
\\
&&
e^{-\gamma(p) \Delta t}
\left(\begin{array}{cc}
        \displaystyle{\cosh{\gamma \Delta t}} &
        \displaystyle{\sinh{\gamma \Delta t}} \\
        \displaystyle{\sinh{\gamma \Delta t}} &
        \displaystyle{\cosh{\gamma \Delta t}}

\end{array}\right)
\left(\begin{array}{c} W(x,p, t) \\ W(x,-p, t) \end{array}\right) 
.
\nonumber
\end{eqnarray}
For sufficiently small time steps $\Delta t$, the combination of these
two steps gives an efficient way to compute the time evaluation of 
the Wigner distribution.

\begin{figure}
\centerline{
\resizebox{8.5cm}{!}{
   \includegraphics*{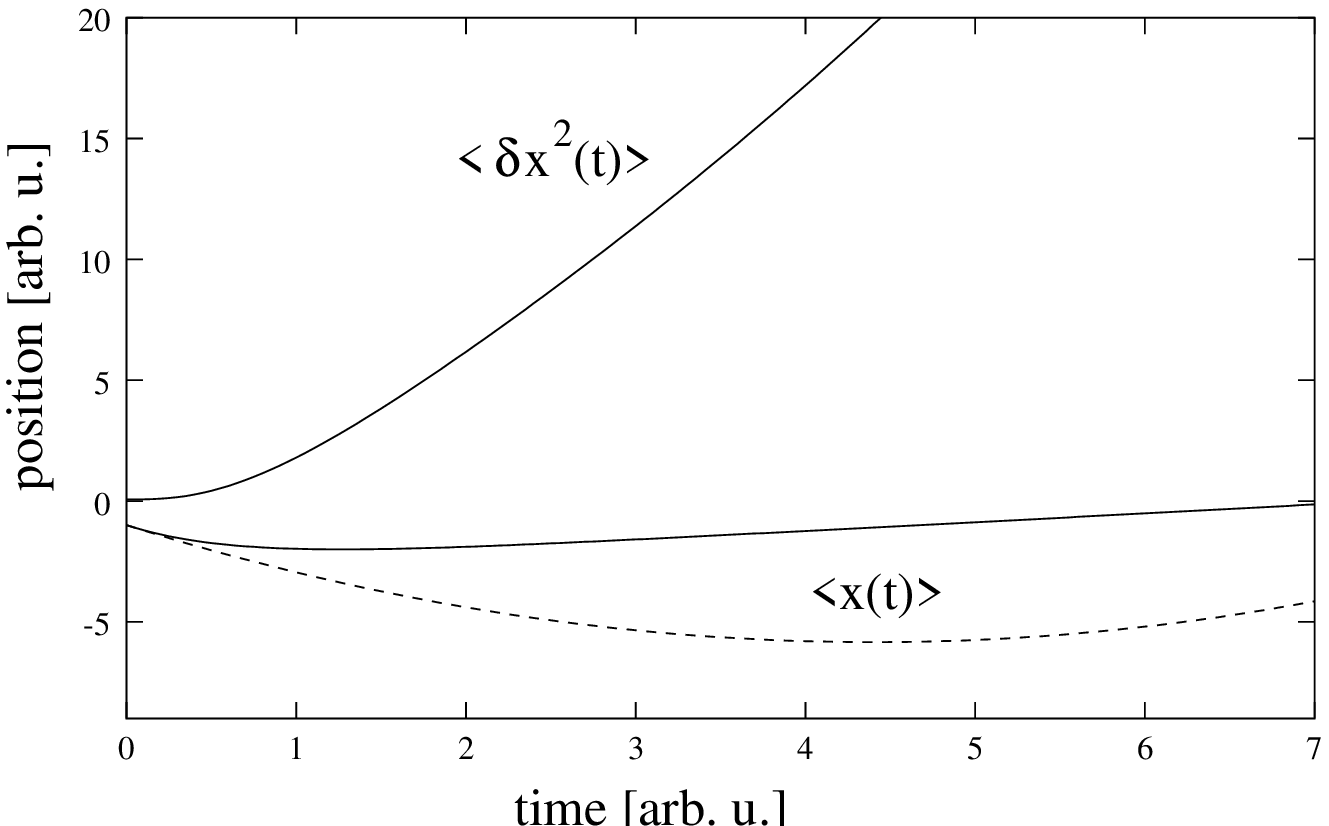}
   \hspace*{10mm}
   \includegraphics*{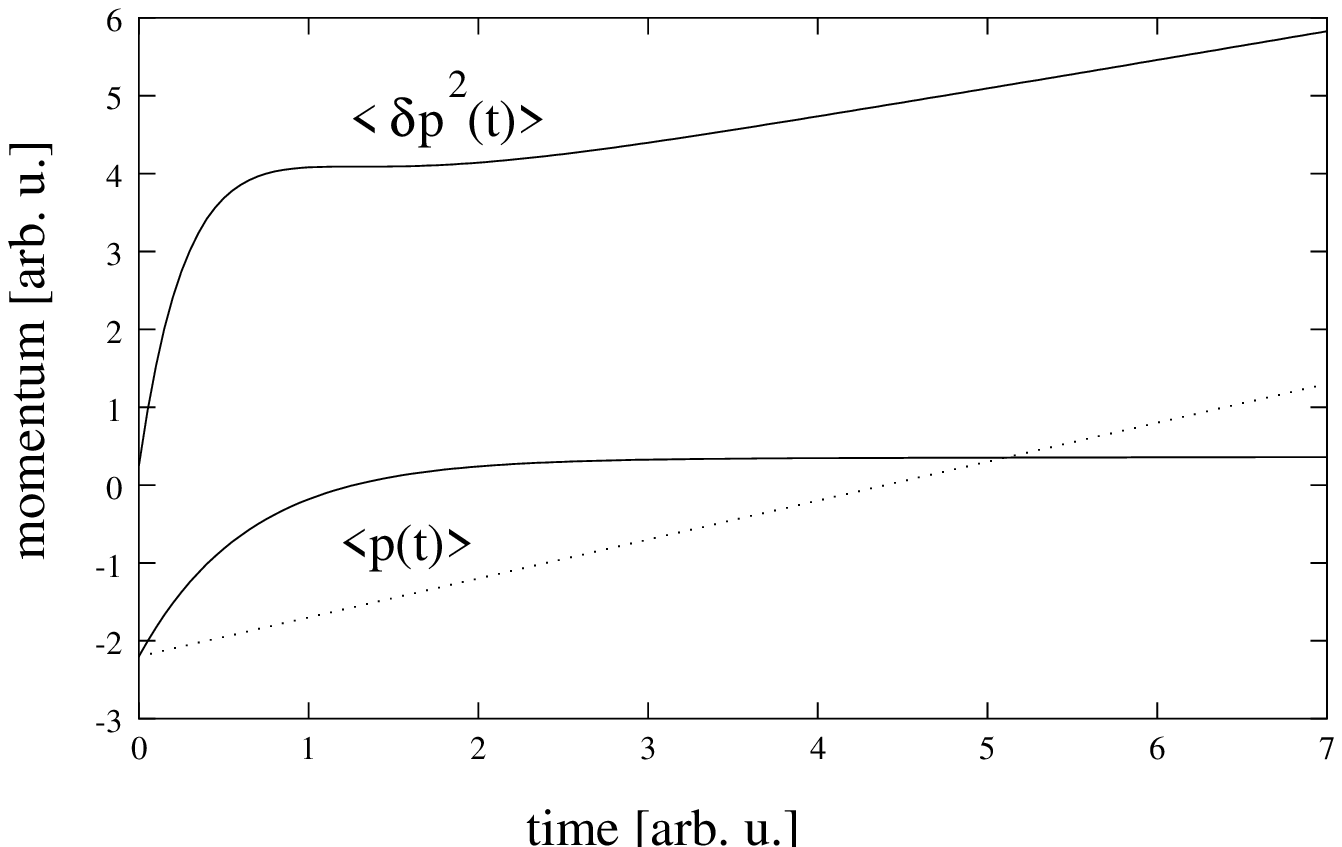}
   }}
\caption[]{
Numerical calculation of the temporal evolution of an atomic cloud in a one-dimensional
waveguide. The cloud is subject to a positive external force
and elastic scattering by a static rough potential, with a negative initial  
average velocity. The two lower curves in the left graph show the evolution of the 
average position $\langle x(t) \rangle$ determined numerically (solid line) and the 
ballistic short time behaviour (dotted line). The upper left curve displays the simulated 
variance $\langle \delta x^2(t) \rangle$ of the width in position space. The two lower 
curves on the right are the numerical 
values (solid line) for the average momentum $\langle p(t) \rangle$ of the cloud 
and the according short time ballistic acceleration (dotted line). The upper curve on the 
right illustrates the numerically computed width in momentum space 
$\langle \delta p^2(t)\rangle$.  
\newline
The scattering rate is taken as $\gamma(p) = \gamma_0 {\rm e}^{ - |p| l_c }$
with $l_c = 0.1\,$units, $\gamma_0 = 1\,$unit.
}
\label{fig:statisticsWithForce}
\end{figure}

\begin{figure}
\centerline{
\resizebox{0.7\columnwidth}{!}{
   \includegraphics*{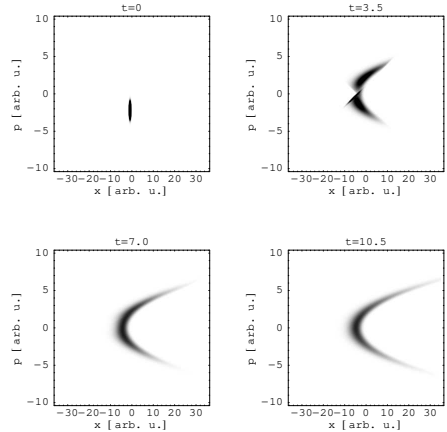}}}
\caption[]{
Snapshots of four stages of the temporal evolution of a Wigner distribution 
accelerated by an external force. The underlying dataset is the same as the one used for 
figure~\ref{fig:statisticsWithForce}. 
}
\label{fig:dynamicsWithForce}
\end{figure}

We present results for the two cases of an atomic sample moving in a wave guide with 
a static rough potential with and without an external force applied that accelerates 
the sample. In the first case we observe that for short times the average motion is 
ballistic, which is shown in figure~\ref{fig:statisticsWithForce}:
the cloud is accelerated according to 
$\langle x(t) \rangle = \langle x(0) \rangle + \langle p(0)/m \rangle t + (F_{\rm ext}/2m) t^2$. 
However, 
at longer times a different behaviour takes over and the average 
position shows a linear drift.
We also observe a monotonic increase with time of the spatial width 
$\langle \delta x^2(t) \rangle$ of the atomic cloud. 
Figure~\ref{fig:statisticsWithForce} shows that this increase 
is approximately linear, as would be the case for a spatial diffusion process. 
The transient ballistic regime may be understood qualitatively by
inspection of the average momentum $\langle p(t) \rangle$ and momentum width 
$\langle \delta p^2(t) \rangle$, 
also shown in figure~\ref{fig:statisticsWithForce}, as well as of the
snapshots of the temporal evolution in phase space (figure~\ref{fig:dynamicsWithForce}):
the elastic scattering transfers atoms from negative velocities to positive velocities 
until the average velocity vanishes. This is accompanied by an increase of the momentum spread.
At the end of the transient regime, both the momentum spread increase slows down and 
the motion in position space is no longer ballistic. Note that the deceleration by the 
scattering potential is faster than expected from the short time acceleration 
$\langle p(t) \rangle = \langle p(0) \rangle + F_{\rm ext}t$. 

For the case of a vanishing external force we now show that
the characteristic behaviour of the atomic cloud can also be 
obtained analytically. The discussion is restricted to $D=1$ for simplicity. 

\subsubsection{Analytical estimate of the diffusion coefficient.}

In the absence of an external force, the momentum $p$ enters the
transport equation only as a parameter, and we are left with a coupled
system for the Wigner functions $W(x, \pm p, t)$ 
[cf.\ also~(\ref{eq:scattering-step})]. Using a spatial Fourier transform
and a temporal Laplace transform (with conjugate variables $k$ and $\zeta$), 
it is straightforward to obtain the following solution
\begin{eqnarray}
&&
\tilde{W}(k, \pm p; \zeta) = 
\\
&& 
\frac{
( \gamma(p) + \zeta - {\rm i} k p/m ) \tilde{W}_0(k, + p)
+
\gamma(p) \tilde{W}_0(k, - p) }{
\zeta^2 + 2 \gamma(p) \zeta + k^2 p^2 / m^2 }
,
\nonumber
\end{eqnarray}
where $\tilde{W}_0(k, + p)$ is the initial (Fourier transformed)
Wigner function. The smallest of the denominator's roots $\zeta_{1,\,2}$ 
determines the long-time behaviour of the Wigner function. For small
values of $k$ (corresponding to large spatial scales), it is given
by
\begin{equation}
\zeta_{\rm min} \approx - \frac{ k^2 p^2 }{ 2 m^2 \gamma(p) }
.
\end{equation}
For a Wigner function with gaussian initial data, we thus find the
following long-time asymptotics:
\begin{eqnarray}
&&
t \to \infty:
\nonumber\\
&&
\langle x(t) \rangle = \langle x(0) \rangle + \frac{ p }{ 2 m \gamma(p) }
,
\label{eq:ave-pos}
\\
&&
\langle \delta x^2(t) \rangle = \langle \delta x^2(0) \rangle 
+ \frac{ p^2 t }{ 2 m^2 \gamma(p) }
.
\label{eq:ave-width}
\end{eqnarray}
The average position thus tends to a constant, that is displaced by
the free flight distance during the scattering time $1/\gamma(p)$.
For longer times, the `forward' and `backward' moving Wigner 
distributions $W(x, \pm p, t)$ have the same weight, and there is
no spatial displacement on average. The spatial width increases
linearly with time, in accordance with a random walk: for each 
scattering time $1/\gamma(p)$, a position step $\sim p/m \gamma(p)$
is performed in a random direction. 

\begin{figure}
\centerline{
\resizebox{0.7\columnwidth}{!}{
   \includegraphics{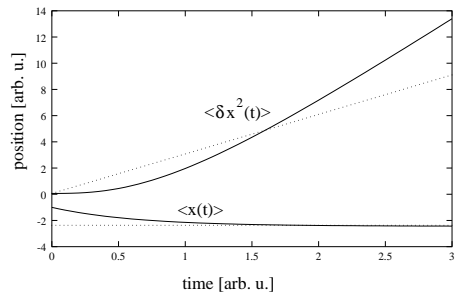}}}
\caption[]{
A free evolving Wigner distribution subject to elastic scattering, 
the distribution having initially a negative average velocity. 
The two lower curves show the evolution of the 
average position $\langle x(t) \rangle$ determined numerically (solid line) and from the 
analytical solution~(\ref{eq:ave-pos}) (dotted line), whereas the upper curves displays 
the numerical average width  $\langle \delta x^2(t) \rangle$ of the distribution (solid line) 
and its analytical estimate from (\ref{eq:ave-width}) (dotted line).
}
\label{fig:statisticsNoForce}
\end{figure}

\begin{figure}
\centerline{
\resizebox{0.7\columnwidth}{!}{
   \includegraphics*{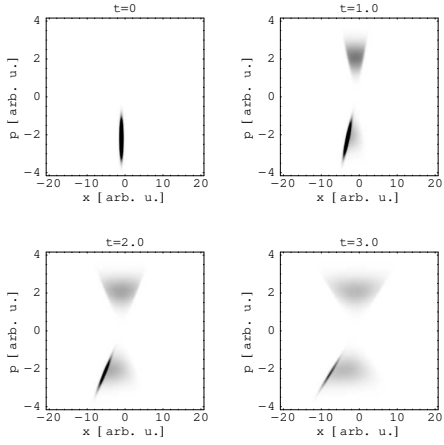}}}
\caption[]{
Snapshots of four stages of the temporal evolution of a Wigner distribution 
only subject to static roughness and vanishing external force. The underlying dataset 
is the same as the one used for figure~\ref{fig:statisticsNoForce} and \ref{fig:coherenceNoForce}. 
}
\label{fig:dynamicsNoForce}
\end{figure}

In figure~\ref{fig:statisticsNoForce}, we compare these results to a numerical solution 
of the transport equation without external force. 
The predictions~(\ref{eq:ave-pos}) for the average position and (\ref{eq:ave-width}) 
for the spatial width show a quite good agreement especially for the average position. 
The transient regime that is responsible for equilibration of the velocity classes
however is not captured by the analytical solution and thus the spatial width is only a
good estimate for the numerical solution, but definitely both exhibit the predicted linear
behaviour. Figure~\ref{fig:dynamicsNoForce} shows the dynamical evolution of the 
Wigner distribution, illustrating the growth of momentum classes with opposite signs.

\begin{figure}
\centerline{
\resizebox{0.7\columnwidth}{!}{
   \includegraphics{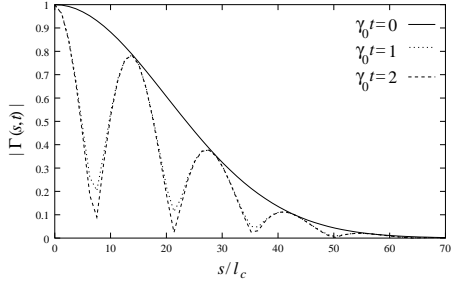}}}
\caption[]{
Illustration of spatial decoherence in an atomic wave guide, caused by
elastic scattering due to static roughness (no external force). 
The spatially averaged coherence function $\Gamma({\bf s}, t)$
is plotted vs.\ the separation ${\bf s}$ for a few times $t$.
Space is scaled to the field correlation length $l_c$ and time
to the scattering time $1/\gamma_0 = 1/\gamma(0)$. The curves were extracted from the
dataset underlying also figure~\ref{fig:statisticsNoForce}. 
}
\label{fig:coherenceNoForce}
\end{figure}

The structure of the Wigner distribution in phase space when the second momentum class 
builds up also modifies the spatially averaged coherence 
function~(\ref{eq:def-coherence-function}) as can be seen in figure~\ref{fig:coherenceNoForce}. 
Since the scattering conserves energy, the momentum width of the cloud and hence the spatial 
coherence length remain constant in time, in distinction
to the inelastic scattering shown in fig.\ref{fig:inelastic-decoh}. The oscillations of the
coherence function are a reminiscence of the standing wave formed by the two velocity classes. 
Note, however, that this interference is `classical' in the sense that the Wigner function 
is positive all over phase space in this case.


\section{Conclusion}
\label{s:conclu}

In this paper, a transport theory for matter wave transport in 
low-dimensional waveguides has been formulated and solved in a few
interesting limiting cases. We have shown that the scattering from
thermal magnetic near fields that `leak out' of metallic microstructures
at room temperature limits the coherence of a cold atomic cloud.
After a few scattering times, the spatial coherence of the ensemble gets 
reduced to the correlation length of the magnetic noise. This length
scale is typically comparable to the distance of the waveguide to the
metallic structures. Our results indicate that decoherence may be
reduced by working with smaller metallic structures, reducing their
temperature and their specific conductivity. Hopefully, a 
reasonable compromise between these conflicting requirements may be found. 

The decoherence rates obtained in the present framework are quite
large and should give rise to measurable effects when interference
experiments with trapped matter waves are performed. From the theoretical
point of view, it is imperative to generalize the transport theory
to Bose-condensed samples: the transport equation then becomes nonlinear,
and the potential fluctuations may expel particles out of the condensate.
This issue is currently under study and results will be reported
elsewhere. Finally, it is well known that in one dimension, static
noise suppresses spatial diffusion, leading to Anderson localisation
\cite{Anderson58}. This effect is beyond the scope of the present
(semiclassical) theory, and its observation in atomic waveguides
would be a clear manifestation of quantum-mechanical interference 
in multiple scattering.

\begin{small}
\paragraph{Acknowledgments.}
We thank F.~Spahn, A.~Pikovsky, and A.~Demircan for interesting exchanges
on the one-dimensional transport equation and S.~A.\ Gardiner 
and A.~Albus for useful discussions on extensions to Bose-condensed
samples. C.~H.\ acknowledges M.~Wilkens for continuous support. 
S.~P.\ was supported in part by the Office of Naval Research
Research Contract No. 14-91-J1205, the National Science Foundation
Grant PHY98-01099, the US Army Research Office and the Joint
Services Optics Program.
\end{small}

\appendix

\section{Low-frequency magnetic noise}
\label{a:noise-currents}

As outlined in subsec.~\ref{s:magnetic-near-field}, we model the
thermal excitations of the metallic structure by currents with a
correlation function
\begin{eqnarray}
&&
\langle 
{j}_i^*( {\bf x}_1; \omega )
{j}_j( {\bf x}_2; \omega' )
\rangle
=
\nonumber\\
&&
4\pi \hbar \varepsilon_0 \omega^2 
\bar n( \omega )
\,
\delta_{ij}
\,
\delta( \omega - \omega' )
{\rm Im}\,\varepsilon( {\bf x}_1 )
\,
\delta( {\bf x}_1 - {\bf x}_2 )
,
\label{eqa:currents}
\end{eqnarray}
where $\bar n( \omega )$ is the Bose-Einstein occupation number. The
current correlation is  $\delta$-correlated
in space provided the dielectric function is local.
In the magnetostatic approximation, the vector potential generated
by the currents is given by (in SI units)
\begin{equation}
{\bf A} ( {\bf x}; \omega ) = 
\frac{ \mu_0 }{ 4 \pi } \int\!{\rm d}^3 x' \,
\frac{
{\bf j}( {\bf x}'; \omega )
}{
|{\bf x} - {\bf x}'|
}
.
\end{equation}
The cross correlation tensor of the vector potential is then simply
obtained from the thermal average of 
${\bf A}_i^*( {\bf x}_1; \omega ) {\bf A}_j( {\bf x}_2; \omega' )$.
Taking the curl with respect to both ${\bf x}_1$ and ${\bf x}_2$, we
get the magnetic cross correlation tensor:
\begin{eqnarray}
&&
\langle B_i^*( {\bf x}_1, \omega )  
B_j( {\bf x}_2, \omega' ) \rangle
= 
\nonumber\\
&& \qquad
=
2\pi \delta( \omega - \omega' )
B_{ij}( {\bf x}_1, {\bf x}_2 ; \omega )
,
\\
&&
B_{ij}( {\bf x}_1, {\bf x}_2 ; \omega ) 
=
S_B^{(bb)}( \omega ) \frac{ 3\, {\rm Im}\,\varepsilon }{ 4\pi \omega / c } 
Y_{ij} 
,
\label{eq:B-quasi-static}
\\
&& Y_{ij} = 
\left(
\delta_{ij} {\rm tr}\,(X_{ij}) - X_{ij} 
\right)
,
\label{eq:X-tensor}
\\
&& 
X_{ij} 
=
\int_V\!{\rm d}^3x' \,
\frac{ 
 ( {\bf x}_1 - {\bf x}' )_i
 ( {\bf x}_2 - {\bf x}' )_j
}{
 | {\bf x}_1 - {\bf x}' |^3
 | {\bf x}_2 - {\bf x}' |^3
}
.
\label{eq:X-tensor-2}
\end{eqnarray}
We have normalized the spectral density to Planck's blackbody formula 
\begin{equation}
S_B^{(bb)} = \frac{ \hbar \omega^3 }{ 3 \pi \varepsilon_0 c^5
( {\rm e}^{ \hbar \omega / k_{\rm B} T } - 1 ) }
.
\end{equation}
The integration in~(\ref{eq:X-tensor-2}) runs over the 
volume $V$ occupied by the thermal
currents (where the imaginary part of $\varepsilon( {\bf x}', \omega )$
is nonzero). Putting ${\bf x}_1 = {\bf x}_2 = {\bf x}$ 
and using the scattering rate~(\ref{eq:flip-rate}), we find 
from~(\ref{eq:B-quasi-static}) that the scattering rate is of the
form~(\ref{eq:general-estimate}). 

For homogeneous metallic structures, the dielectric function is
$\varepsilon( {\bf x}', \omega ) = {\rm i}/( \varepsilon_0 \varrho \omega)$
where $\varrho$ is the {\sc dc} conductivity. In addition,  
the relevant frequencies are so low that the high-temperature limit
of the Planck function is applicable. 
The prefactor in~(\ref{eq:B-quasi-static}) then approaches the constant value
\begin{equation}
\omega\to 0: \quad
S_B^{(bb)}( \omega ) \frac{ 3\, {\rm Im}\,\varepsilon
}{ 
4\pi \omega / c } 
\to
\frac{ k_{\rm B} T }{ 4\pi^2 \varepsilon_0^2 c^4 \varrho }
,
\label{eq:low-freq-constant}
\end{equation}
and we find the prefactor in~(\ref{eq:low-freq-limit}). We have compared the results
of the magnetostatic approximation to exact calculations using the retarded Green 
function for a layered medium. Both calculations agree in the short-distance limit 
where the distances are small compared to both the wavelength and the skin depth 
$\delta = \sqrt{ 2 \varepsilon_0 \varrho / (\omega c^2)}$. However, 
the magnetostatic approximation overestimates the field components parallel 
to the layer by a factor of three. This is related to the fact that the noise 
currents~(\ref{eqa:currents}) are not divergence-free at the surface. Possible
improvements will be discussed elsewhere in more detail.

\clearpage

\end{document}